\documentclass[preprint,showpacs,preprintnumbers,amsmath,amssymb,floatfix,endfloats*]{revtex4}

\usepackage{graphicx}
\usepackage{dcolumn}
\usepackage{bm}
\usepackage{amsfonts}

\DeclareMathAlphabet{\mathsfsl}{OT1}{cmr}{bx}{it}
\begin{document}
\title{Nonaffine rearrangements of atoms in deformed and quiescent binary glasses}
\author{Nikolai V. Priezjev}
\affiliation{Department of Mechanical and Materials Engineering,
Wright State University, Dayton, OH 45435}
\date{\today}
\begin{abstract}

The influence of periodic shear deformation on nonaffine atomic
displacements in an amorphous solid is examined via molecular
dynamics simulations. We study the three-dimensional Kob-Andersen
binary mixture model at a finite temperature.  It is found that when
the material is periodically strained, most of the atoms undergo
repetitive nonaffine displacements with amplitudes that are broadly
distributed.   We show that particles with large amplitudes of
nonaffine displacements are organized into compact clusters.   With
increasing strain amplitude, spatial correlations of nonaffine
displacements become increasingly long-ranged, although they remain
present even in a quiescent system due to thermal fluctuations.

\end{abstract}




\maketitle

\section{Introduction}

Elucidating the connection between atomic structure and mechanical
properties of amorphous materials is important for a variety of
optical and structural applications~\cite{FalkRev16}.       Unlike
crystalline solids, where plasticity is determined by motion of
topological line defects called dislocations, an elementary process
that leads to macroscopic plastic deformation of amorphous solids
involves a collective rearrangement of a small number of atoms,
often referred to as a shear transformation
zone~\cite{Argon79,Falk98}. Hence, the localized plastic activity in
a deformed material is associated with large nonaffine displacements
of atoms, i.e., when atomic displacements do not match the
macroscopic strain. Surprisingly, it was found that spatial
correlations of nonaffine displacements in steadily sheared
amorphous solids are long-ranged~\cite{Chikkadi12,Varnik14}.
Recently, many experiments have been conducted  to study
shear-induced rearrangements of atoms and the yielding transition in
glassy materials subjected to oscillatory shear
strain~\cite{Dennin08,Pusey08,Arratia13,Cipelletti14,Echoes14,Ganapathy14,Zaccone15,Zaccone16}.
However, the nature of spatial correlations of plasticity in glasses
under periodic shear deformation remains not fully understood.

\vskip 0.05in

In the last few years, a number of studies have investigated the
mechanical response of amorphous solids to oscillatory shear strain
using atomistic quasistatic simulations in the athermal
limit~\cite{Reichhardt13,Sastry13,HernHoy13,IdoNature15}.   It was
found that at small strain amplitudes, the systems evolve into
periodic limit cycles where particles follow the same trajectories
over consecutive cycles, and the number of cycles to reach a limit
cycle increases as a critical strain amplitude is approached from
below~\cite{Reichhardt13,IdoNature15}.     Interestingly, reversible
avalanches of large particle displacements were observed at strain
amplitudes below the critical value~\cite{IdoNature15}. At finite
temperatures, it was shown that the mean square displacement of
particles exhibits a broad subdiffusive plateau during ten thousand
cycles even at small strain amplitudes, and the relaxation process
involves intermittent clusters of particles undergoing cage
jumps~\cite{Priezjev13,Priezjev14}.     It was also found that at
the critical strain amplitude, the number of mobile particles
involved in a correlated motion reaches
maximum~\cite{Priezjev13,Priezjev14}.

\vskip 0.05in

The shear transformation zones were directly observed in recent
experiments on colloidal glasses under reversible macroscopic shear
deformation and at mechanical equilibrium~\cite{Spaepen14}.     The
spatial correlations of the local strain field clearly showed the
fourfold patterns, characteristic of Eshelby inclusions, that are
present even in a quiescent system where their orientation is
isotropic, thus producing zero global strain~\cite{Spaepen14}. In
turn, a local reversible shear transformation in a quiescent glass
can trigger irreversible cage jumps whose density is large when the
system dynamics is weakly damped or the shear transformation is
slow~\cite{Priezjev15,Priezjev15a}. More recently, a sharp
transition from affine to nonaffine displacements of particles was
identified in colloidal glasses under variable-amplitude oscillatory
shear~\cite{Zaccone15,Zaccone16}. It was recently shown that during
periodic deformation of binary glasses, particles undergo repetitive
nonaffine displacements whose amplitudes become more broadly
distributed with increasing strain amplitude~\cite{Priezjev16}.
However, the spatial organization and correlations of nonaffine
displacements as well as their occurrence in a quiescent system were
not explored in detail.

\vskip 0.05in

In this paper, molecular dynamics simulations are performed to study
nonaffine atomic displacements in a three-dimensional binary mixture
model subjected to slow periodic shear at a finite temperature.  We
show that the oscillatory deformation induces repetitive collective
rearrangements of atoms that deviate significantly from the affine
displacement field at large strain amplitudes.    It is found that
in a deformed material, the spatial correlations of nonaffine
displacements are long-ranged, while they become short-ranged in a
quiescent system, where nonaffine rearrangements are thermally
activated.

\vskip 0.05in

The remainder of the paper is organized as follows. The details of
molecular dynamics simulations are given in the next section.   The
numerical results for the oscillatory shear deformation and
quiescent systems and the analysis of nonaffine rearrangements of
atoms are presented in Sec.\,\ref{sec:Results}.    The summary is
given in the last section.

\section{Molecular dynamics simulations}
\label{sec:MD_Model}

\begin{figure}[t]
\includegraphics[width=10.cm,angle=0]{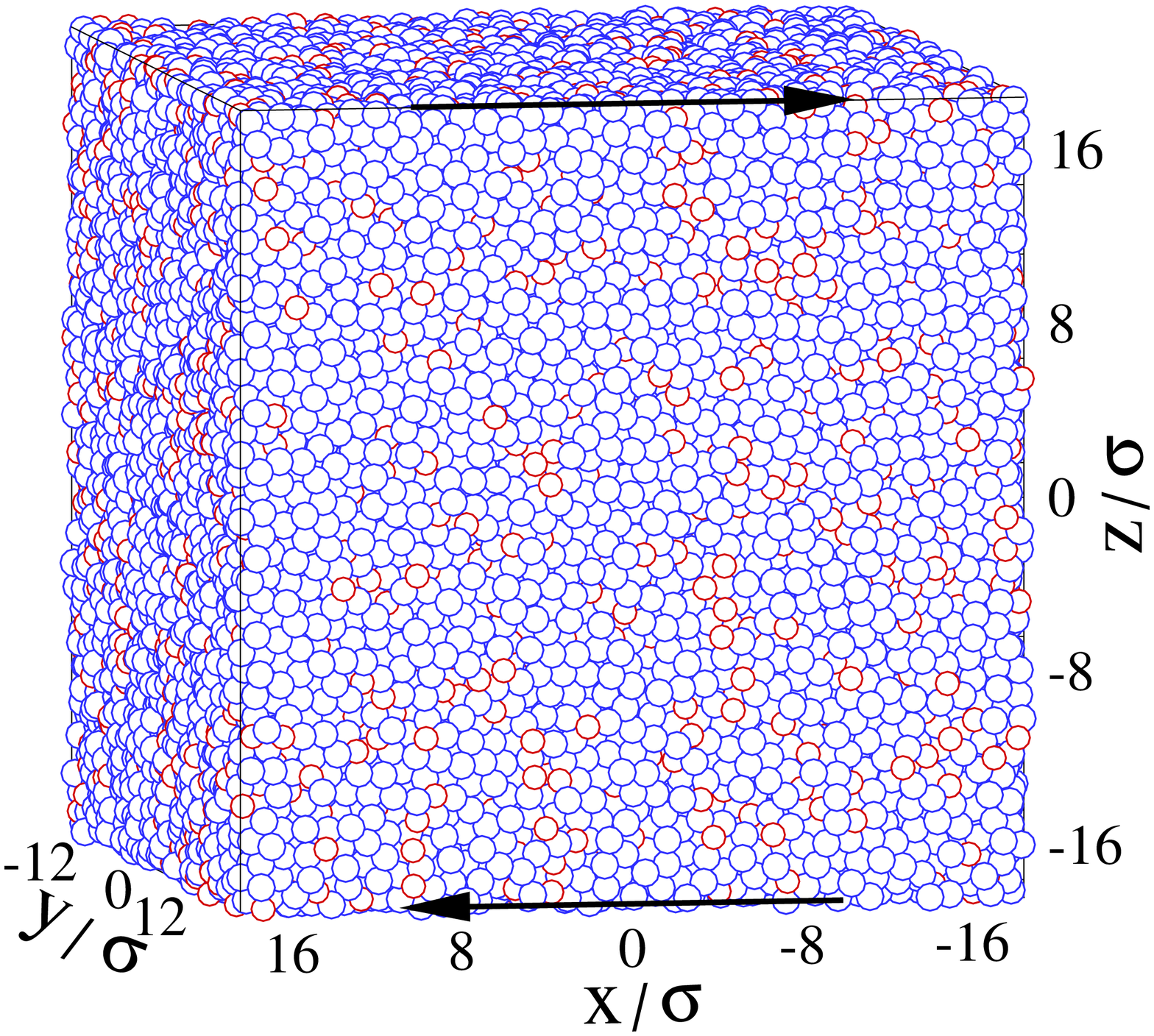}
\caption{(Color online) The instantaneous atomic positions for a
binary LJ glass annealed with a slow cooling rate of
$10^{-5}\,\varepsilon/k_B\tau$ to the temperature
$T_{LJ}=10^{-2}\,\varepsilon/k_B$.    Atoms of type A are shown by
large blue circles and type B by small red circles.    The arrows
indicate the plane of periodic shear deformation.  }
\label{fig:snapshot_system}
\end{figure}

The molecular dynamics simulations were performed using the
open-source LAMMPS numerical code developed at Sandia National
Laboratories~\cite{Lammps}.    We use the standard Kob-Andersen
model~\cite{KobAnd95} of a binary Lennard-Jones glass.   Our
three-dimensional system consists of $N=60\,000$ atoms confined in a
periodic box as shown in Fig.\,\ref{fig:snapshot_system}. Two types
of atoms are considered, $\alpha,\beta=A,B$, which interact via the
truncated LJ potential as follows:
\begin{equation}
V_{\alpha\beta}(r)=4\,\varepsilon_{\alpha\beta}\,\Big[\Big(\frac{\sigma_{\alpha\beta}}{r}\Big)^{12}\!-
\Big(\frac{\sigma_{\alpha\beta}}{r}\Big)^{6}\,\Big],
\label{Eq:LJ_KA}
\end{equation}
where the parameters are set to $\varepsilon_{AA}=1.0$,
$\varepsilon_{AB}=1.5$, $\varepsilon_{BB}=0.5$, $\sigma_{AB}=0.8$,
$\sigma_{BB}=0.88$, and $m_{A}=m_{B}$~\cite{KobAnd95}.    The cutoff
radius is fixed to $r_{c,\,\alpha\beta}=2.5\,\sigma_{\alpha\beta}$.
As usual, the units of length, mass, energy are set
$\sigma=\sigma_{AA}$, $m=m_{A}$, $\varepsilon=\varepsilon_{AA}$,
and, correspondingly, the unit of time is
$\tau=\sigma\sqrt{m/\varepsilon}$.    The equations of motion were
integrated using the Verlet algorithm~\cite{Allen87} with the time
step $\triangle t_{MD}=0.005\,\tau$.

\vskip 0.05in


The atoms were initially arranged at the sites of the face-centered
cubic lattice within a cubic box of linear dimension
$L=36.84\,\sigma$.     Throughout the study, the density of atoms
was kept constant $\rho=\rho_{A}+\rho_{B}=1.2\,\sigma^{-3}$. First,
the system was equilibrated in the absence of deformation at the
high temperature $1.1\,\varepsilon/k_B$, which is well above the
computer glass transition temperature
$T_g\approx0.45\,\varepsilon/k_B$~\cite{KobAnd95}.   Here $k_B$
denotes the Boltzmann constant.     The temperature was controlled
by the Nos\'{e}-Hoover thermostat with the damping time of
$1.0\,\tau$. Then, the system was gradually cooled with a
computationally slow rate of $10^{-5}\,\varepsilon/k_B\tau$ to the
target temperature $T_{LJ}=10^{-2}\,\varepsilon/k_B$.  This
procedure was repeated for twenty independent samples.

\vskip 0.05in


The periodic shear strain deformation of the material was applied in
the $xz$ plane using the Lees-Edwards periodic boundary
conditions~\cite{Allen87}, while the system remained periodic along
the $\hat{y}$ direction.  The shear strain was varied as a function
of time according to
\begin{equation}
\gamma(t)=\gamma_{0}\,\,\textrm{sin}(2\pi t / T),
\label{Eq:strain}
\end{equation}
where $\gamma_{0}$ is the strain amplitude and $T=10^4\tau$ is the
oscillation period.   The corresponding oscillation frequency is
$\omega=2\pi/T=6.28 \times 10^{-4}\,\tau^{-1}$.  We note that the
streaming velocity was subtracted from the local particle velocity
to compute temperature in the Nos\'{e}-Hoover
thermostat~\cite{Lammps}.  The system was subject to periodic
deformation during $10$ cycles and the positions of all atoms were
saved every $T/100=100\,\tau$. The analysis of nonaffine
displacements was performed in twenty independent samples in the
range of strain amplitudes $0 \leqslant \gamma_{0} \leqslant 0.08$
during the last five cycles.

\section{Results}
\label{sec:Results}


The time-periodic deformation of an amorphous solid typically
involves an alternating sequence of elastic strain separated by
rapid plastic events, which are accompanied by sudden drops in
stress and the potential energy~\cite{IdoNature15}.   Depending on
the strain amplitude, the local rearrangements of particles can be
reversible after one or more cycles or irreversible leading to
chaotic system dynamics and particle diffusion~\cite{IdoNature15}.
In our study, the periodic shear strain is spatially homogeneous,
which prevents the formation of shear bands during slow, nearly
quasistatic, deformation period $T=10^4\tau$.

\vskip 0.05in

\begin{figure}[t]
\includegraphics[width=12.cm,angle=0]{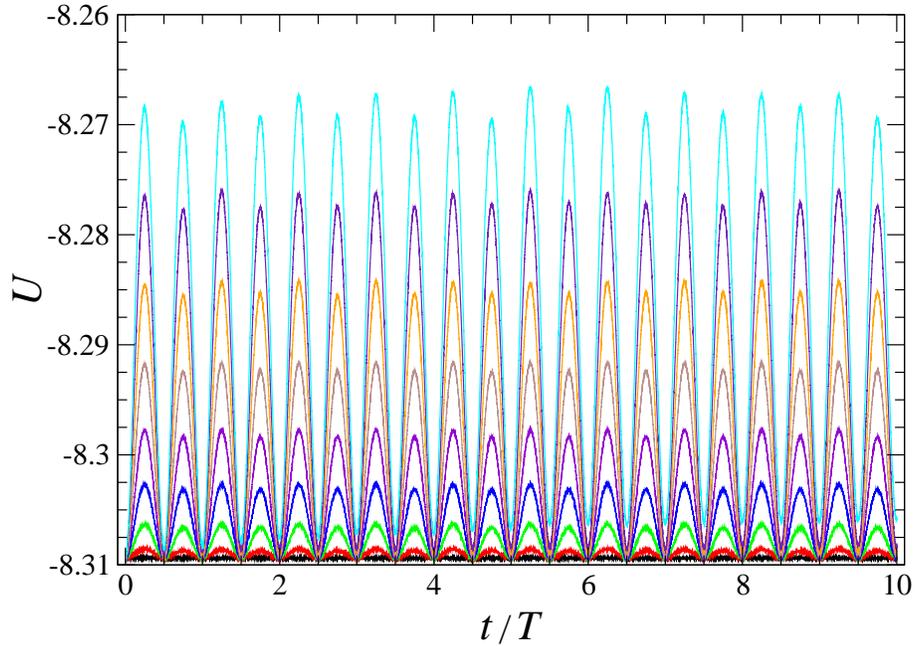}
\caption{(Color online) The potential energy per atom $U$ (in units
of $\varepsilon$) during ten oscillation cycles for the strain
amplitudes $\gamma_{0} = 0$, $0.01$, $0.02$, $0.03$, $0.04$, $0.05$,
$0.06$, $0.07$, $0.08$ (from bottom to top). }
\label{fig:poten}
\end{figure}

The potential energy per atom is plotted in Fig.\,\ref{fig:poten}
for one representative sample, which was deformed during ten
back-and-forth cycles at different strain amplitudes.   The
horizontal black line at $U\approx -8.31\,\varepsilon$ corresponds
to the potential energy measured in the quiescent system.   Note
that for each curve, the superimposed noise is caused by thermal
motion of atoms.   At small strain amplitudes $\gamma_{0} < 0.06$,
the minima of the potential energy at the end of each cycle remain
unchanged, whereas at large strain amplitudes $\gamma_{0} \geqslant
0.06$, the potential energy increases over consecutive cycles. These
results imply that during periodic shear deformation at small strain
amplitudes, the system dynamics is reversible after each cycle. In
contrast, with increasing strain amplitude, $\gamma_{0} \geqslant
0.06$, some atoms undergo irreversible displacements leading to a
progressive increase in the potential energy.

\vskip 0.05in


These conclusions are consistent with the energy landscape picture
of sheared glasses; namely, that a cycle of large shear strain
rejuvenates the amorphous material by relocating the system to
shallower energy minima~\cite{Lacks04}.   Depending on the annealing
procedure, it was also shown that a cycle with a small strain
amplitude overages the glass by moving the system to deeper energy
minima~\cite{Lacks04}.   In our simulations, the decrease in the
potential energy after several deformation cycles is not observed at
small strain amplitudes because of the computationally slow cooling
rate that brings the system to an initial state with a very low
energy minimum.

\vskip 0.05in


In general, during an affine deformation of the material, the
displacement of atoms can be described by a combination of a linear
transformation and a translation.  Correspondingly, a deviation from
the linear strain produces a finite nonaffine component of
displacement of atoms with respect to their neighbors. The measure
of the nonaffine displacement of the atom $\mathbf{r}_{i}(t)$ during
the time interval $\Delta t$ is defined as follows:
\begin{equation}
D^2(t, \Delta t)=\frac{1}{N_i}\sum_{j=1}^{N_i}\Big\{
\mathbf{r}_{j}(t+\Delta t)-\mathbf{r}_{i}(t+\Delta t)-\mathbf{J}_i
\big[ \mathbf{r}_{j}(t) - \mathbf{r}_{i}(t)    \big] \Big\}^2,
\label{Eq:D2min}
\end{equation}
where the sum is taken over $N_i$ nearest-neighbor atoms within the
distance $1.5\,\sigma$ from $\mathbf{r}_{i}(t)$, and $\mathbf{J}_i$
is the transformation matrix that best maps all bonds between the
$i$-th atom and its nearest-neighbors at times $t$ and $t+\Delta
t$~\cite{Falk98,Ma12}.   In the present study, the quantity $D^2(t,
\Delta t)$ was computed every $100\,\tau$ during each cycle with
respect to zero global strain.


\vskip 0.05in

\begin{figure}[t]
\includegraphics[width=12.cm,angle=0]{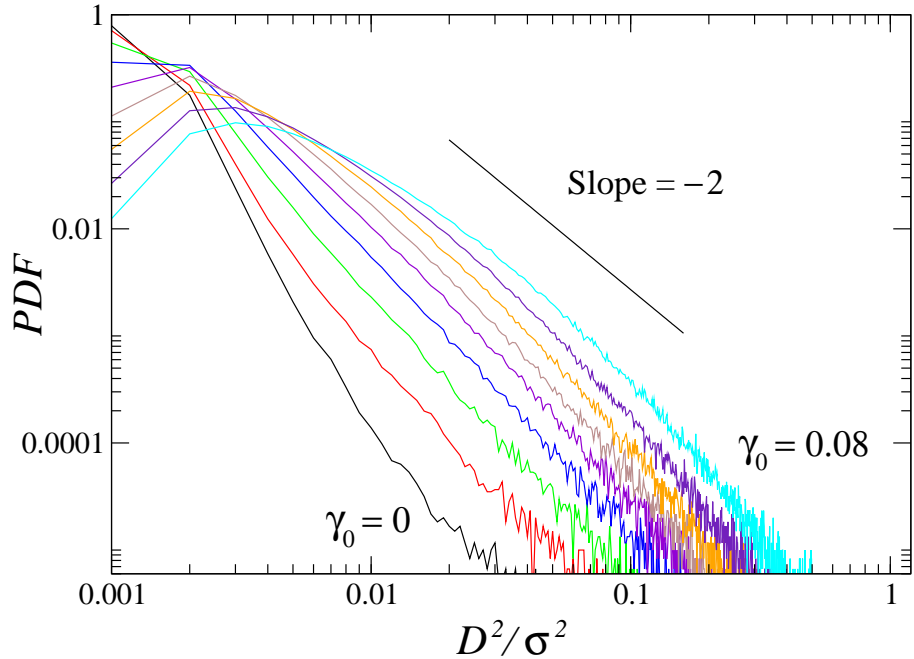}
\caption{(Color online) The probability distribution function of
$D^2(t, T/4)$ for the strain amplitudes $\gamma_{0} = 0$, $0.01$,
$0.02$, $0.03$, $0.04$, $0.05$, $0.06$, $0.07$, $0.08$ (from left to
right).  The straight solid line indicates the slope $-2$. }
\label{fig:d2min_pdf}
\end{figure}

In driven amorphous systems at finite temperatures, nonaffine
rearrangements of atoms arise due to both external stresses and
thermal fluctuations~\cite{Liu15}.  The normalized probability
distribution function of nonaffine displacements is shown in
Fig.\,\ref{fig:d2min_pdf} for various strain amplitudes.   The
quantity $D^2(t, \Delta t)$ was averaged over all atoms in twenty
systems at the maximum strain $\Delta t = T/4$ with respect to zero
strain.    It can be seen that in the quiescent system, when
$\gamma_{0}=0$, nonaffine deformations are present but their
distribution is relatively narrow and bounded by
$D^2(T/4)\approx0.03\,\sigma^2$, which is comparable to the cage
size $r_{cage}\approx0.1\,\sigma$. It should be noted that during
the time interval $T/4$, the motion of nearest neighbors is
uncorrelated, and, therefore, the quantity $D^2(T/4)$ essentially
measures the average nonaffine rearrangement between two random
configurations of atoms within their cages.  To remind, in a dense
amorphous system at the low temperature
$T_{LJ}=10^{-2}\,\varepsilon/k_B$, most of the atoms in the absence
of external deformation remain within their cages, or escape only
temporarily, on the time scale accessible to computer
simulations~\cite{KobAnd95}.

\vskip 0.05in


With increasing strain amplitude, the probability distribution
becomes more skewed towards larger values of $D^2(T/4)$ with the
power-law exponent approaching $-2$ at the largest strain amplitude
$\gamma_{0}=0.08$, as shown in Fig.\,\ref{fig:d2min_pdf}.   The
nonaffine displacements of atoms with $D^2(T/4)\gg0.01\,\sigma^2$
reflect that their nearest-neighbor structure is significantly
deformed during periodic strain. Moreover, the power-law
distribution of $D^2(T/4)$ implies that highly nonaffine
displacements are collectively organized, and, as will be shown
below, they tend to form compact clusters.  We further comment that
the distribution functions shown in Fig.\,\ref{fig:d2min_pdf} change
continuously near the critical strain amplitude $\gamma_{0}=0.06$,
which marks the transition from a slow relaxation dynamics with a
subdiffusive plateau to a diffusive
regime~\cite{Priezjev13,Priezjev16}. Finally, it was also recently
reported that the probability distribution function of $D^2(\Delta
t)$, as well as other nonaffine measures based on displacement
fluctuations and on deviation from the global deformation, exhibit a
power-law decay with the slope $-2.8$ in steadily sheared colloidal
glasses~\cite{Chikkadi12}.

\vskip 0.05in

\begin{figure}[t]
\includegraphics[width=12.cm,angle=0]{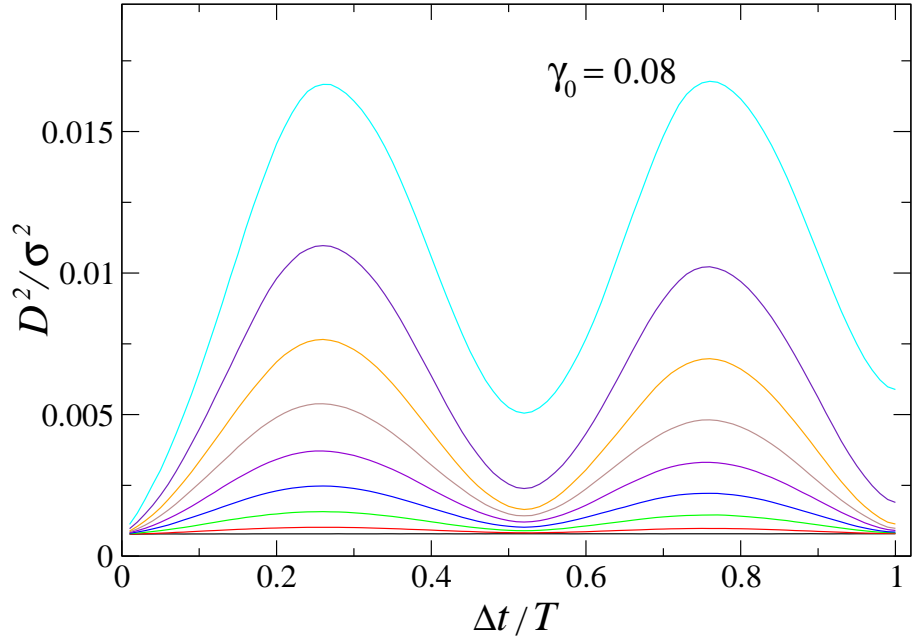}
\caption{(Color online) Variation of the averaged quantity $D^2(t,
\Delta t)$ during one oscillation period for the strain amplitudes
$\gamma_{0} = 0$, $0.01$, $0.02$, $0.03$, $0.04$, $0.05$, $0.06$,
$0.07$, $0.08$ (from bottom to top). }
\label{fig:d2min_over_T}
\end{figure}

Next, we plot the variation of the quantity $D^2(t, \Delta t)$ in
Fig.\,\ref{fig:d2min_over_T} during one oscillation period for the
indicated strain amplitudes.   The data were averaged over all atoms
during the last five cycles.    It can be observed that in the
absence of external deformation, $D^2(t, \Delta t)$ is finite and
time-independent.    As mentioned above, nonaffine displacements in
a quiescent system arise due to thermal vibration of atoms within
their cages, and their local configurations become uncorrelated over
time intervals $\Delta t \geqslant 0.01\,T=100\,\tau$.    Since the
mean square displacement of atoms within their cages is proportional
to temperature, it is expected that $D^2(\Delta t)$ scales as
$T_{LJ}$. Furthermore, as the strain amplitude increases, nonaffine
displacements become more pronounced, especially near the maximum
strain at $\Delta t = T/4$ and $3\,T/4$ (see
Fig.\,\ref{fig:d2min_over_T}).    More importantly, the quantity
$D^2(T)$ after a full cycle is greater than $D^2(0.01\,T)$ at large
strain amplitudes $\gamma_{0} \geqslant 0.06$, which implies that
some atoms undergo irreversible displacements during one cycle. In
marked contrast, at small strain amplitudes $\gamma_{0} < 0.06$, the
nonaffine measure $D^2(0.01\,T) \approx D^2(T)$, indicating
reversible dynamics.    This transition correlates well with the
increase in the potential energy over consecutive cycles at
$\gamma_{0} \geqslant 0.06$ as shown in Fig.\,\ref{fig:poten}.

\vskip 0.05in

\begin{figure}[t]
\includegraphics[width=12.cm,angle=0]{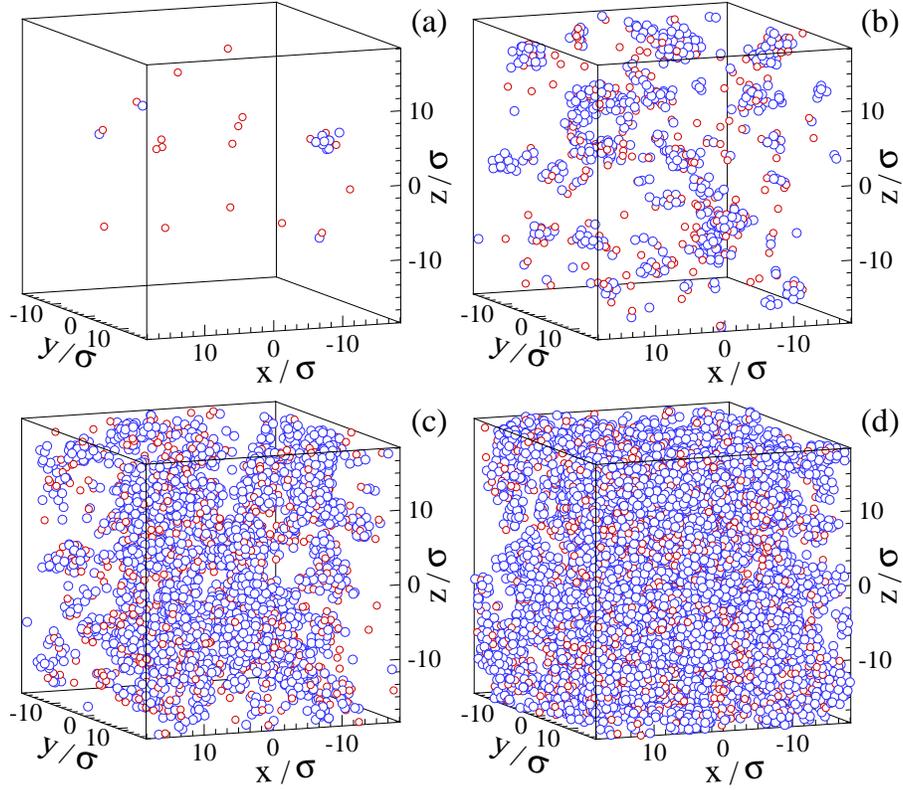}
\caption{(Color online) Spatial configurations of atoms with large
nonaffine displacements after a quarter of a cycle, $D^2(t,
T/4)>0.01\,\sigma^2$, for the strain amplitudes (a)
$\gamma_{0}=0.00$, (b) $\gamma_{0}=0.02$, (c) $\gamma_{0}=0.04$, and
(d) $\gamma_{0}=0.06$.   The atoms of type A and B are denoted by
blue and red circles. }
\label{fig:snapshot_clusters}
\end{figure}

Figure\,\ref{fig:snapshot_clusters} illustrates the spatial
organization of atoms with large nonaffine displacements, $D^2(t,
T/4)>0.01\,\sigma^2$, for selected strain amplitudes.   Notice that
in the quiescent system, there are only several isolated atoms,
mostly smaller atoms of type B, with large nonaffine components;
although one group of atoms forms a small cluster [see
Fig.\,\ref{fig:snapshot_clusters}\,(a)].    At the strain amplitude
$\gamma_{0} = 0.02$, localized regions of highly nonaffine
displacements can be observed in
Fig.\,\ref{fig:snapshot_clusters}\,(b).  At larger strain
amplitudes, $\gamma_{0} = 0.04$ and $0.06$, the clusters become
interconnected and comparable with the system size.     The
appearance of repetitive clusters of atoms with large nonaffine
displacements is consistent with reversible avalanches of particles
rearrangements, which are associated with large drops in the
potential energy, during periodic deformation of an athermal solid
below the critical strain amplitude~\cite{IdoNature15}.    Moreover,
it was also shown that energy drops are approximately power-law
distributed in a 2D solid under oscillatory shear below the yielding
transition~\cite{IdoNature15}.    In our simulations, the cluster
size distribution was not computed due to insufficient statistics,
since only several large clusters are typically present in each
independent sample~\cite{Priezjev16}.

\vskip 0.05in

\begin{figure}[t]
\includegraphics[width=12.cm,angle=0]{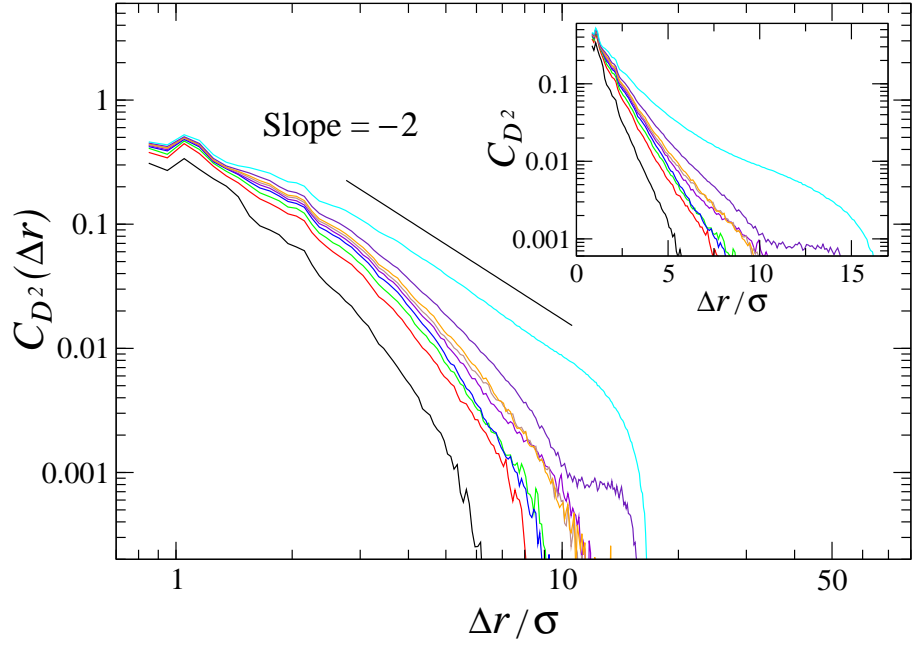}
\caption{(Color online) The correlation function $C_{D^2}(\Delta
\textbf{r})$ defined by Eq.\,(\ref{Eq:CORR_D2}) for the strain
amplitudes $\gamma_{0} = 0$, $0.01$, $0.02$, $0.03$, $0.04$, $0.05$,
$0.06$, $0.07$, $0.08$ (from left to right).   The straight line
with the slope $-2$ is plotted for reference. The inset shows the
same data on the log-normal scale. }
\label{fig:corr_d2min}
\end{figure}

A related analysis of the spatial correlations of nonaffine
displacements, however, can be performed by considering the
normalized, equal-time correlation
function~\cite{Chikkadi12,Varnik14,Murali12} given by
\begin{equation}
C_{D^2}(\Delta \textbf{r}) = \frac{\langle D^2(\textbf{r} + \Delta
\textbf{r}) D^2(\textbf{r}) \rangle - \langle D^2(\textbf{r})
\rangle^2}{\langle D^2(\textbf{r})^2 \rangle - \langle
D^2(\textbf{r}) \rangle^2},
\label{Eq:CORR_D2}
\end{equation}
where the brackets denote averaging over all pairs of atoms in
twenty independent samples, and the quantity $D^2$ is computed at
$\Delta t = T/4$ with respect to zero strain.   The correlation
function $C_{D^2}(\Delta \textbf{r})$ is shown in
Fig.\,\ref{fig:corr_d2min} for quiescent and periodically strained
glasses.    It is clearly seen that with increasing strain
amplitude, the spatial correlations become stronger and more
long-ranged.   At the largest strain amplitude $\gamma_{0} = 0.08$,
the data can be well described by a power-law decay with the
exponent $-2$, followed by a cutoff due to the finite system size.
As shown in the inset of Fig.\,\ref{fig:corr_d2min}, the decay of
$C_{D^2}(\Delta \textbf{r})$ at $\gamma_{0} = 0$ is exponential,
which indicates that correlations of thermally induced nonaffine
rearrangements in a quiescent glass extend up to nearest neighbor
distances. Recently, it was pointed out that in steadily sheared
glasses, spatial correlations of $D^2(\Delta \textbf{r})$ at
intermediate times change from an exponential to a power-law decay
with the exponent of about $-1.3$ when the system size is
increased~\cite{Varnik14}.   In the present study, most of the
rearrangements are reversible after one deformation cycle in a
relatively large system; and the crossover from the exponential to
the power-law decay of nonaffinity correlations occurs due to the
increase in the strain amplitude.

\section{Conclusions}

Using molecular dynamics simulations of a binary Lennard-Jones glass
at a finite temperature, we have investigated the spatial
distribution and correlation of plasticity in periodically strained
and quiescent systems.  In a three-dimensional amorphous glass, the
nonaffine displacements were computed for each atom after multiple
time intervals with respect to zero macroscopic strain.  It was
shown that nonaffine displacements arise both in quiescent and
periodically deformed glasses. At mechanical equilibrium, thermal
motion of atoms within their cages leads to a relatively narrow
distribution of nonaffine displacements and their spatial
correlations extend up to nearest neighbor distances. During
periodic deformation at small strain amplitudes, nonaffine
displacements are shear induced, spatially heterogeneous, and
reversible after each cycle. Above the critical strain amplitude,
some atoms undergo irreversible nonaffine rearrangements and the
potential energy of the system increases over consecutive cycles.
With increasing strain amplitude, nonaffine displacements become
more broadly distributed and organized into large interconnected
clusters.  The numerical results indicate that spatial correlations
of nonaffinity are approximately power-law distributed at large
strain amplitudes.

\vskip 0.05in

In the future, it will be instructive to extend this study to larger
systems and longer times in order to evaluate more accurately the
cluster size distribution of nonaffine displacements and their
spatial correlations, and to compare these results with experiments
on periodically deformed colloidal glasses or bulk metallic glasses.

\section*{Acknowledgments}

Financial support from the National Science Foundation (CNS-1531923)
is gratefully acknowledged.  The molecular dynamics simulations were
conducted using the LAMMPS numerical code~\cite{Lammps}.
Computational work in support of this research was performed at
Michigan State University's High Performance Computing Facility and
the Ohio Supercomputer Center.



\bibliographystyle{prsty}

\end{document}